\documentclass[11pt,a4paper]{article}
\usepackage{amssymb}
\usepackage{amsmath}\usepackage{graphicx}  
\numberwithin{equation}{section}
\usepackage{footnote}
\makesavenoteenv{tabular}

\newcommand{\id}[1]{\ensuremath{\mathrm{id}}}

\newcommand{\qm}{quantum mechanics}

\newcommand{\beq}{\begin{equation}}
\newcommand{\eeq}{\end{equation}} 
\newcommand{\bea}{\begin{eqnarray}}
\newcommand{\eea}{\end{eqnarray}}

\newcommand{\la}{\langle} \newcommand{\ra}{\rangle}

\newcommand{\tp}{transition probability}
\newcommand{\tpies}{transition probabilities}

\newcommand{\ca}{C*-algebra} 
 \newcommand{\rep}{representation}

\newcommand{\Hs}{Hilbert space}

\newcommand{\lm}{\lambda} 
\newcommand{\rh}{\rho} 
  
 \newcommand{\phv}{\varphi}
\newcommand{\ch}{\ch}

\newcommand{\Tr}{\mbox{\rm Tr}\,}

 \newcommand{\CP}{{\mathcal P}}

\newcommand{\C}{{\mathbb C}} 
 \newcommand{\R}{{\mathbb R}}
 
  \makeatletter
 
\makeatletter
\def\moverlay{\mathpalette\mov@rlay}
\def\mov@rlay#1#2{\leavevmode\vtop{%
   \baselineskip\z@skip \lineskiplimit-\maxdimen
   \ialign{\hfil$\m@th#1##$\hfil\cr#2\crcr}}}
\newcommand{\charfusion}[3][\mathord]{
    #1{\ifx#1\mathop\vphantom{#2}\fi
        \mathpalette\mov@rlay{#2\cr#3}
      }
    \ifx#1\mathop\expandafter\displaylimits\fi}
\makeatother

\topmargin = - 1 cm \textheight = 23 cm \textwidth = 15 cm
\begin{document} 
\pagenumbering{arabic} \setlength{\unitlength}{1cm}\cleardoublepage
\date\nodate
\renewcommand{\thefootnote}{\fnsymbol{footnote}}
\begin{center}
\begin{huge}
{\bf  Quantum theory and functional analysis}\footnote{To appear in the \emph{Oxford Handbook of the History of Interpretations and Foundations of Quantum Mechanics}, ed.\ O. Freire (Oxford University Press, 2021). This chapter suffers from a strict word limit, as a consequence of which the discussion is often terse. For example, instead of explaining the technical details, for which I refer to books like Landsman (2017), I have tried to sketch the relevant history at an almost sociological level. 
I am deeply indebted to Michel Janssen and Miklos R\'{e}dei for helpful comments.}
\end{huge}
\bigskip

\bigskip

\begin{Large}
 Klaas Landsman
 \end{Large}
 \bigskip
 
 \begin{large}
Institute for Mathematics, Astrophysics, and Particle Physics (IMAPP), \\
Faculty of Science, Radboud University, Nijmegen, The Netherlands

 and 
 
 Dutch Institute for Emergent Phenomena (DIEP), \texttt{www.d-iep.org}.

Email:
\texttt{landsman@math.ru.nl}
 \end{large}
\bigskip

\end{center}
 \begin{abstract} 
 \noindent
Quantum theory and functional analysis were created and put into essentially their final form during similar periods ending around 1930. Each was also a key outcome of the major revolutions that both physics and mathematics as a whole underwent at the time. This paper studies their interaction in this light, emphasizing the leading roles played by Hilbert in preparing the ground and by von Neumann in  bringing them together during the crucial year of 1927, when he gave the modern, abstract definition of a Hilbert space and applied this concept to quantum mechanics (consolidated in his famous monograph from 1932). 
Subsequently, I give a very brief overview of three areas of
 functional analysis that have had fruitful interactions with quantum theory since 1932,
namely unbounded operators, operator algebras, and distributions. The paper closes with some musings about the  role of functional analysis in actual physics. 
  \end{abstract}
\bigskip
\tableofcontents

\thispagestyle{empty}
\renewcommand{\thefootnote}{\arabic{footnote}}
\newpage \setcounter{footnote}{0}
\section{Introduction}
Dijksterhuis (1961) concludes his masterpiece  \emph{The Mechanization of the World Picture} (which ends with Newton)  with the statement that the process described in the title consisted of the mathematization of the natural sciences, adding that this process had been completed by twentieth-century physics. As such, the topic of this chapter seems a perfect illustration of  Dijksterhuis's claim, perhaps even \emph{the} most perfect illustration.\footnote{Einstein's theory of General Relativity is an equally deep and significant example of the process in question, but I would suggest that his application of Riemannian geometry to physics was, though unquestionably an all-time highlight of  science, less unexpected than the application of functional analysis to quantum theory. Indeed,  Riemann certainly thought about field theory and gravity in this connection.} 

However, there is an important difference between the application of calculus to classical mechanics and  the application of functional analysis to quantum mechanics: Newton invented calculus in the context of classical mechanics,\footnote{The fact that Newton subsequently erased his own calculus from the \emph{Principia} does not change this.} whereas functional analysis was certainly \emph{not} created with quantum theory in mind. In fact, the interaction between the two fields only started in 1927, when quantum mechanics was almost finished at least from a physical point of view, and also functional analysis had most of its history behind it.\footnote{See  Bernkopf (1966), Monna (1973), Steen (1973), Dieudonn\'{e} (1981), Birkhoff and Kreyszig (1984), Pier (2001), and Siegmund-Schultze (1982, 2003) 
 for the historical development of functional analysis. According to most authors this history occupied a period of about 50 years, starting in the 1880s and ending in 1932 with the books by Banach, Stone, and von Neumann published in that year (see below).}

Functional analysis \emph{did} have its roots in \emph{classical} physics. Monna (1973), Dieudonn\'{e} (1981), and  Siegmund-Schultze (2003)  trace functional analysis  back to various sources:
\begin{enumerate}
  \item \emph{The Calculus of Variations}, which by itself has a distinguished history involving J. Bernoulli, Euler, Lagrange, Legendre, Jacobi, and others. This was one of the sources of  the idea of studying spaces of functions (though not necessarily \emph{linear} ones) and functionals (\emph{idem dito}) thereon, that is, ``functions of functions". 
This was picked up in the 1880s by the so-called Italian school of functional analysis, involving Ascoli and Arzel\`{a} (whose theorem was the first rigorous result in the subject), 
Volterra, Pincherle, and to some extent Peano (1888), who first axiomatized linear spaces.\footnote{See Monna (1973), Dorier (1995) and Moore (1995)
 for the history of linear structures.}  
\item \emph{Infinite systems of linear equations with an infinite number of unknowns}, initially coming from Fourier's work on heat transfer in the 1820s; his idea of what we now call ``Fourier analysis" links certain linear partial differential equations (PDEs) with such systems. This link was almost immediately generalized by the 
  \emph{Sturm--Liouville theory} of linear second-order differential equations from the 1830s, in which the crucial idea of eigenfunctions and eigenvalues originates.
Around 1890,  the analysis of Hill, Poincar\'{e}, and von Koch on the motion of the moon provided further inspiration.
\item \emph{Integral equations}, first studied by Abel in the 1820s and independently by Liouville in the 1830s in connection with problems in mechanics. From the 1860s onwards integral equations  were used by Beer, Neumann, and others as a tool in the study of harmonic functions and the closely related \emph{Dirichlet problem} (which asks for a function satisfying Laplace's equation on a domain with prescribed boundary value, and may also be seen as a variational problem). This problem, in turn,
 came from the study of vibrating membranes via PDEs from the 18th century onwards.\footnote{It is hardly a coincidence that the
 Dirichlet problem was eventually solved rigorously in 1901 by none other than Hilbert (Monna, 1975), whose 
 role in functional analysis is described below and in \S\ref{HFA} .}
\end{enumerate}
On the other hand, functional analysis  benefited from--and was eventually even one of the highlights of--the abstract or ``modernist"  turn that mathematics took in the 19th century.\footnote{See Mehrtens (1990) and Gray (2008) in general, and Siegmund-Schultze (1982) for functional analysis. } In my view (supported by what follows), it was exactly this turn that made the completely unexpected application of functional analysis to quantum theory possible, and hence it seems no accident that 
Hilbert was a crucial player both in the decisive phase of the modernist turn \emph{and} in the said application.  Thus Hilbert 
 played a double role in this:
\begin{enumerate}
\item Through his general views on mathematics  (which of course he instilled in his pupils such as Weyl and von Neumann) and the ensuing scientific atmosphere he had created in G\"{o}ttingen.\footnote{`\emph{One cannot overstate the significance of the influence exerted by Hilbert's thought and personality on all who came out of  [the Mathematical Institute at G\"{o}ttingen]}' (Corry, 2018). See also Rowe (2018). }
 Hilbert's views branched  off in two closely related directions:
\begin{itemize}
\item His relentless emphasis on \emph{axiomatization}, which started (at least in public) with his famous memoir \emph{Grundlagen der Geometry} from 1899 (Volkert, 2015).
\item His promotion of the interplay between mathematics and  physics (Corry, 2004a).
\end{itemize}
These came together in his  \emph{Sixth Problem} (from the famous list of 23 in 1900):\footnote{See e.g.\ Corry (1997, 2004a, 2018) and references therein. It is puzzling that Hilbert did not mention Newton's \emph{Principia} in this light, which was surely the first explicit and successful axiomatization of physics.}
\begin{quote}\begin{small}
Mathematical Treatment of the Axioms of Physics. The investigations on the foundations of geometry suggest the problem: To treat in the same manner, by means of axioms, those physical sciences in which already today mathematics plays an important part; in the first rank are the theory of probabilities and mechanics. (Hilbert, 1902).\end{small}
\end{quote}
\item Through his contributions to functional analysis, of which he was one of the founders:
\begin{quote}\begin{small}
By the depth and novelty of its ideas, [Hilbert (1906)]  is a turning point in the history of Functional Analysis, and indeed deserves to be considered the very \emph{first} paper published in that discipline. (Dieudonn\'{e}, 1981, p.\ 110). \end{small}
\end{quote}
\end{enumerate}
From both an intellectual  and an institutional point of view, the connection between quantum theory and functional analysis could be made (at least so quickly) because in the 1920s G\"{o}ttingen did not only have the best mathematical institute in the world (with a tradition going back to Gau\ss, Riemann, and now Hilbert), but, due to the presence of Born, Heisenberg, and Jordan, and others, was also one of the main centers in the creation of \qm\  in the crucial years 1925--1927. It was this combination that enabled the decisive contributions of von Neumann (who spent 1926--1927  in  G\"{o}ttingen, see \S\ref{vN}).

It cannot be overemphasized how remarkable the link between quantum theory and functional analysis is. 
The former is the physical theory of the atomic world that was developed between 1900--1930, written down for the first time in systematic form in Dirac's celebrated textbook  \emph{The Principles of Quantum Mechanics} from 1930 (Jammer, 1989). The latter is a mathematical theory of infinite-dimensional vector spaces (and linear maps between these), endowed with some notion of convergence (i.e.\ a topology), either in abstract form or in concrete examples where the ``points" of the space are often functions.  

These two topics appear to have nothing to do with each other whatsoever, and hence the work of von Neumann (1932) in which they are related seems nothing short of a miracle. The aim of my paper is to put this miracle in some historical perspective.
 \newpage
\section{Hilbert: Axiomatic method}\label{HAM}
\begin{quote}
\begin{small} I believe this: as soon as it is ripe for theory building, anything that can be the subject of scientific thought at all 
 falls under the scope of the axiomatic method and hence indirectly of mathematics.
 By penetrating into ever deeper layers of axioms in the sense outlined erlier we also gain insight into the nature of scientific thought by itself and become steadily more aware of the unity of our knowledge. Under the header of the axiomatic method mathematics appears to be called into a leading role in science in general. (Hilbert, 1918, p.\ 115).\end{small}
 \end{quote}
 Hilbert  (1918) begins his essay on axiomatic thought (of which the above text is the end) by stressing the importance of the connection between mathematics and neigbouring fields, especially physics and epistemology, and then says that the essence of this connection lies in the \emph{axiomatic method}. By this, he simply means the identification of certain sentences (playing the role of axioms) that form the foundation of a specific field in the sense that its theoretical structure (Hilbert uses the German word \emph{Fachwerk}) can be (re)constructed from the axioms via logical principles. Axioms  typically state relations between ``things" (\emph{Dinge}), like ``points" or ``lines", which are defined implicitly through the axioms and hence may change their meaning  if the axiom systems in which they occur change, as is the case in e.g.\ non-Euclidean geometry.\footnote{The revolutionary nature of this view may be traced from Hilbert's correspondence with Frege, who apparently never accepted  (or even grasped) this point (Gabriel \emph{et al}, 1980; Blanchette, 2018).}
 The epistemological status of the axioms differs between fields. For example, Hilbert considered geometry initially a natural science:
 \begin{quote}\begin{small}
Geometry also emerges from the observation of nature, from experience. To this extent, it is an \emph{experimental science}. (\ldots) All that is needed is to derive [its] foundations from a minimal set of \emph{independent axioms} and thus to construct the whole edifice of geometry by \emph{purely logical means}. In this way geometry is turned into a \emph{purely mathematical science}.\end{small} (Hilbert in 1898--99, quoted in Corry, 2004a, p.\ 90).\footnote{From unpublished lecture notes by Hilbert, emphasis in original. Translation:  Corry.}
\end{quote}
This does not mean that he treated the axioms of geometry  as ``true" (as Euclid had done): Hilbert often stressed the tentative and malleable nature of axiom systems,\footnote{As exemplified by the seven editions of \emph{Grundlagen der Geometrie}  Hilbert published during his lifetime!} and acknowledged that axioms for physics might even be inconsistent, in which case finding new, consistent axioms is an important source of progress (Corry, 2004a; Majer, 2014).\footnote{Though Einstein would speak of ``principles" rather than ``axioms", many of his key contributions to physics, such as special relativity, general relativity, and the EPR-argument are examples of this strategy.} 

Hilbert is famous for his purely formal  treatment of axioms,\footnote{This has earned Hilbert the undeserved reputation of being a ``formalist", which is remote from his actual views on mathematics. The  purely symbolic treatment of axioms and proofs was not even new with Hilbert; he apparently took it from Russell  (Mancosu, 2003; Ewald \& Sieg, 2013) and hence indirectly from Frege and Peano. But unlike Russell, until the last decade of his career dedicated to proof theory, Hilbert stated axioms quite informally, using a  combination of mathematical and natural language. } 
 which indeed was striking all the way from the \emph{Grundlagen der Geometrie} in 1899 to his swan song  \emph{Grundlagen der Mathematik} (Hilbert \& Bernays, 1934, 1939), but in fact such formality is always strictly limited to the
 logical analysis of axiom systems (notably his relentless emphasis on consistency and to a lesser extent on completeness)
 and the validation of proofs. Indeed, except for logic Hilbert made almost no contribution to the axiomatization of mathematical structures, although, starting already in the 19th century with e.g.\ Dedekind, Peano and Weber, this became a central driving force of 20th century mathematics (Corry, 2004b).
\section{Hilbert: Functional analysis}\label{HFA}
Inspired by Fredholm's theory of integral operators,\footnote{[The day (in 1901) on which Holmgren spoke on Fredholm's work in Hilbert's seminar] `\emph{was decisive for a long period in Hilbert's life and for a considerable part of his fame}' (Blumenthal, 1935, p.\ 410).} in the paper mentioned by Dieudonn\'{e} in the Introduction above, Hilbert (1906)
 introduced many of the key tools of functional analysis, such as bounded and compact operators and spectral theory, culminating in his discovery of continuous spectra.\footnote{Since he lacked the concept of a linear operator Hilbert used a somewhat cumbersome definition of a spectrum; the modern definition is due to Riesz (1913) and was also adopted by von Neumann, see \S\ref{vN}.}
 However, what we now see as \emph{the} central aspect of functional analysis, namely its linear structure, is absent! Hilbert's analysis is entirely given in terms of quadratic forms $K(x)=\sum_{p,q} k_{pq}x_px_q$, where the sequence $(x)$ satsifies $\sum_k |x|_k^2\leq 1$, so that he works on what we would now call the closed unit ball of the Hilbert space $\ell^2$ of square-summable sequences (which is compact in what we now call the weak topology, which Hilbert also introduced himself and heavily exploited). 
 
 As pointed out at the end of \S\ref{HAM}, though at first sight odd,\footnote{ Dieudonn\'{e} (1981) explains  the 19th century emphasis on matrices and quadratic forms at the expense of vectors and linear maps, so that Hilbert had one foot in the 19th century and the other in the 20th.} it seems typical for Hilbert not to rely on \emph{abstract} axiomatized mathematical structures, let alone that he would care to refer to Peano (1888),  in which the concepts of a vector space and a linear map had been axiomatized. Perhaps Peano's axiomatization was really unknown in G\"{o}ttingen, where Hilbert's former student Weyl (1918) rediscovered the axioms for a vector space in the context of Einstein's theory of General Relativity (at least, he did not cite Peano either).\footnote{See, however, Corry (2004a, \S 9.2) on the culture of ``nostrification" in Hilbert's G\"{o}ttingen: `\emph{It was widely understood, among German mathematicians at least, that ``nostrification" encapsulated the peculiar style of creating and developing scientific ideas in  G\"{o}ttingen, and not least because of the pervasive influence of Hilbert. Of course, ``nostrification" should not be understood as mere plagiarism.}' ({\em loc.\ cit.} p.\ 419). }  However, the essentially linear nature of Hilbert's constructions was soon noted and developed by various mathematicians, notably Hilbert's own student Schmidt (1908), who introduced $\ell^2$ including its inner product and even norm in modern form,\footnote{The first   author to use the term ``Hilbert space" (\emph{Hilbertscher Raum}) was
Sch\"{o}nflies (1908), but he meant the closed unit ball in $\ell^2$ (which was historically spot on, since that is what Hilbert analysed!). Riesz (1913)  used \emph{l'espace hilbertien} for what we now call $\ell^2$, and both the notion and the name \emph{Hilbert space} for the general abstract concept we now take it to mean was introduced by von Neumann (1927a). }
and a bit later by Riesz (1913), who rewrote most of Hilbert's results in the modern way via bounded or compact linear operators on Schmidt's space $\ell^2$ (the notion of a linear operator as such had already appeared before, notably in the Italian school of functional analysis). 

Around 1905, Hadamard and his student Fr\'{e}chet (partly inspired by the Italian school) emphasised the idea of  looking at functions as points in some (infinite-dimensional) vector space, including an early use of topology, then also a new field--it was  Fr\'{e}chet (1906) who  in his thesis introduced metric spaces. This idea, often seen as the essence of functional analysis, crossed the Hilbert school through the introduction of $L^2$-spaces, including the spectacular and unexpected isomorphism $L^2([a,b])\cong\ell^2$ due to Riesz (1907) and Fischer (1907). A truly geometric or spatial view of functional analysis was subsequently developed especially by  Riesz (1913, 1918), culminating in the axiomatic development of Banach spaces in the 1920s by Helly, Wiener, and Banach (Monna, 1973; Pietsch, 2007).\footnote{
As an intermediate step from Hilbert to Banach spaces, $L^p$ spaces were introduced by Riesz (1909). Many historians point out that (Frigyes) Riesz was familiar with the Italian, French, and German schools of functional analysis. 
Dieudonn\'{e} (1981, p.\ 145) calls Riesz (1918), which  develops the (spectral) theory of compact operators on Banach spaces \emph{avant la lettre}, `one of the most beautiful papers ever written.'
} 
\section{von Neumann:  Foundations of quantum theory}\label{vN}
 \emph{Methoden der mathematischen Physik} by Courant and Hilbert (1924) put the lid on the G\"{o}ttingen school in functional analysis. It was meant to mathematize classical physics.
\begin{quote}
\begin{small}
And now, one of those events happened, unforeseeable by the wildest imagination, the like of which could tempt one to believe in a pre-established harmony between physical nature and mathematical mind:\footnote{``Pre-established harmony", a philosophical concept originally going back to Leibniz, was a popular concept in the G\"{o}ttingen of Hilbert, where it referred to the relationship between mathematics and physics, or more generally between the human mind and nature (Pyenson, 1982; Corry, 2004a). Minkowski, Hilbert, Born, and Weyl himself all used it as approriate, and Corry (2004a, pp.\ 393--394) even claims that it was `\emph{one of the most basic concepts that underlay the whole scientific enterprise in G\"{o}ttingen}', adding that `\emph{Hilbert, like all his colleagues in  G\"{o}ttingen, was never really able to explain, in coherent philosophical terms, its meaning and the possible basis of its putative pervasiveness, except by alluding to ``a miracle".}' } Twenty years after Hilbert's investigations \emph{quantum mechanics} found that the observables of a physical system are represented by the linear symmetric operators on a Hilbert space and that the eigen-values and eigen-vectors of that operator which represent \emph{energy} are the energy levels and corresponding stationary quantum states of the system. Of course, this quantum-physical interpretation added greatly to the interest in the theory and led to a more scrupulous investigation of it, resulting in various simplifications and extension.
 (Weyl, 1951, p.\ 541). \end{small}
\end{quote}
Although Weyl (who more often fell into lyrical overstatements in his philosophical writings) may have been right about the events in question being unforeseeable, the historical record shows considerable continuity, too. Perhaps a slightly more balanced judgement is:
\begin{quote}
\begin{small} This revolution was made possible by \emph{combining} a concern for rigorous foundations with an interest in physical applications, \emph{and} by coordinating the relevant literature in depth.
\end{small} (Birkhoff \& Kreyszig, 1984, pp.\ 306--307).
\end{quote}
At least the first two aspects were exemplified by Hilbert, who had lectured on the mathematical foundations of physics since 1898 (Sauer\& Majer, 2009; Majer \& Sauer, 2021), and, helped by various assistants,\footnote{Hilbert's first assistant (at the time unpaid) had been Born in 1904;  from 1922--1926 it was Nordheim. }  organised a regular research seminar on the latest developments in physics (Reid, 1970; Schirrmacher, 2019).  With
 Born, Heisenberg, and Jordan all at  G\"{o}ttingen at the time, during the Winter Semester of 1926--1927 Hilbert lectured on quantum theory, with book-length lecture notes by Nordheim.\footnote{These may be found in Sauer\& Majer (2009), pp.\ 507--706. Half of the course, on the ``old quantum theory" was practically reproduced from Hilbert's earlier lectures on quantum theory  during 1922--1923.} These lectures are very impressive and cover almost everything, from Hamilton--Jacobi theory and the ``old quantum theory" to Heisenberg's matrix mechanics, Schr\"{o}dinger's wave mechanics, Born's probability interpretation, and finally Jordan's \emph{Neue Begr\"{u}ndung}, i.e.\ his attempt (simultaneous with Dirac's) to unify the last three ingredients into a single formalism.\footnote{See Duncan \& Janssen (2009, 2013) for a detailed survey of Jordan's \emph{Neue Begr\"{u}ndung}. Older and somewhat complementary histories of this period include Jammer (1989) and  Mehra \& Rechenberg (2000).}
 
 `Coordinating the relevant literature in depth' was therefore certainly taken care of on the physics side, but on the mathematical side Hilbert's surprising lack of interest in the axiomatisation of new mathematical structures except logic
(cf.\ \S\ref{HAM}) still played a role:
\begin{quote}
\begin{small}
The German school [in functional analysis, i.e.\ Hilbert's school] remained reserved with respect to the more abstract concepts of set theory and axiomatics until well into the 1920s. (Siegmund-Schultze, 2003, p.\ 385).\end{small}
\end{quote}
By a simple twist of fate, in 1926 Hilbert attracted the internationally acknowledged young genius von Neumann to spend  the academic year 1926--1927 to G\"{o}ttingen in order to work on his Proof Theory,\footnote{Hilbert got a fellowship for von Neumann from the International Education Board (a subsidiary of the Rockefeller Foundation). In the Fall of 1927 von Neumann moved to Berlin as a \emph{Privatdocent}, where Hilbert's former student Schmidt provided him with the concept of self-adjointness he had initially missed in setting up his spectral theory for closed unbounded operators (Birkhoff \& Kreyszig, 1984, p.\ 309).}
 but in actual fact the latter mostly worked on the mathematical foundations of quantum theory and thus filled in the abstraction and axiomatisation gap.\footnote{von Neumann had made brilliant contributions to set theory already in his late teens. For further information about von Neumann see  Oxtoby {\em et al} (1958), Heims (1980), 
 Glimm  {\em et al} (1990), 
 Macrae (1992),  Br\'{o}dy \&  V\'{a}mos (1995),
R\'{e}dei (2005b), and the rare but  insightful manuscript Vonneumann (1987).} 

The paper by Hilbert, von Neumann, \& Nordheim (1927), now obsolete, is actually a summary of Hilbert's (inconclusive) views, based on his lectures; the decisive establishment of the interaction between quantum theory and functional analysis is entirely due to  von Neumann (1927ab), with mathematical details further elaborated in  von Neumann (1930ab) and a unified exposition in his famous book from 1932, which remains a classic.
Apart from his discussion of quantum statistical mechanics and of the measurement problem, which were path-breaking contributions to physics but are less relevant for our topic, the main accomplishments of von Neumann (1932) in the light of functional analysis are:\footnote{Even finding \emph{one} of these would have been impressive, not only for someone who was 23 years old.}
\begin{enumerate}
\item Axiomatisation of the notion of a \Hs\ (previously known only in examples).
\item Establishment of a spectral theorem for (possibly unbounded) self-adjoint operators.
\item Axiomatisation of quantum mechanics in terms of Hilbert spaces (and operators):
\begin{enumerate}
\item  Identification of observables with (possibly unbounded) self-adjoint operators.
\item Identification of pure states with one-dimensional projections (or rays).
\item  Identification of transition amplitudes with inner products.
\item A  formula for the Born rule stating the probability of measurement outcomes.
\item Identification of general states with density operators.
\item  Identification of propositions with  closed subspaces (or the projections thereon).
\end{enumerate}
\end{enumerate}
In particular, von Neumann provided \emph{two} separate (but closely related) axiomatisations:\footnote{See 
 Lacki (2000),
R\'{e}dei (2005a), and R\'{e}dei \& St\"{o}ltzner (2006) on von Neumann's methodology.} 
\begin{itemize}
\item of Hilbert space as an abstract mathematical structure (almost 
 \emph{contra} Hilbert);
\item
of \qm\ as a physical theory (entirely  in the spirit of Hilbert).
\end{itemize}
\begin{trivlist}
\item[\emph{Ad 1}.] Although Schmidt, Riesz, and others, had thought about sequence and function spaces like $\ell^2$ and $L^2$
in a geometric way 20 years earlier, including the use of inner products, orthogonality, and norms, the abstract concept of a \Hs\ (unlike that of a Banach space) was still lacking before 1927.  The novelty of von Neumann's coordinate-free approach to \Hs s is illustrated by the fatherly advice Schmidt gave him: 
\begin{quote}
\begin{small}
No! No! You shouldn't say operator,  say matrix!
 (Bernkopf, 1967, p.\ 346).\end{small}
\end{quote}
 \item[\emph{Ad 2}.] This was a vast abstraction and generalisation of practically all of the spectral theory done in Hilbert's school, including the work of Weyl and Carleman on what (since von Neumann) are called unbounded self-adjoint operators and their deficiency indices.\footnote{This history is very nicely explained by Dieudonn\'{e} (1981), Chapter VII. See also Stone (1932).}
\item[\emph{Ad 3}.] This remains the basis for any  discussion of the foundations of quantum theory.
\item[\emph{Ad 3(a)}.] One of von Neumann's main goals was a rigorous proof of the equivalence between matrix mechanics (which almost deliberately lacked states) and wave mechanics  (which initially lacked observables). As a first ingredient,
 Heisenberg's matrices (which by themselves were `quantum-mechanical reinterpretations of classical observables') were reinterpreted once again, now as self-adjoint operators on some \Hs\ like $\ell^2$. The need for \emph{unbounded} operators, e.g.\ for position, momentum, and energy, emerged at once.
\item[\emph{Ad 3(b)}.] And this was the second ingredient of the equivalence proof. Identifying Schr\"{o}ding\-er's wave-function $\Psi$ with a unit vector in the \Hs\ $L^2(\R^3)$ was an accomplishment by itself, but on top of this, von Neumann (and Weyl) quickly recognized the importance of the fact that such vectors only define (pure)  states \emph{up to a phase}. Thus the cleanest way to define (pure) states is to identify them with $1d$ projections rather than vectors.\footnote{
Following Minkowski's example from his Geometry of Numbers, in the completely different context of \qm\ von Neumann defined pure states to be extreme points of the convex set of all states. }
\item[\emph{Ad 3(c)}.] Next to the goal just stated, \emph{the} point of von Neumann's axiomatisation was to provide a home to the mysterious transition amplitudes $\la\phv|\psi\ra$
 that were at the heart of Jordan's \emph{Neue Begr\"{u}ndung} (Duncan \& Janssen, 2013), which also Born, Dirac, and Pauli regarded as the essence of \qm. If $|\psi\ra$ and $|\phv\ra$ are unit vectors in some \Hs, von Neumann took the amplitude $\la\phv|\psi\ra$ to be their inner product, with corresponding \tp\ $P(\phv,\psi)=|\la\phv| \psi\ra|^2$. In terms of the one-dimensional projections $e=|\phv\ra\la\phv|$ and $f=|\psi\ra\la\psi|$, this gives
$P(e,f)=\Tr(ef)$, where  $\mathrm{Tr}$ is the trace.\footnote{
It should be admitted that this did not  clinch the issue. Dirac and Jordan  used 
probability amplitudes like $\la x|p\ra=\exp(-ixp/\hbar)$, where $x,p\in\R$, but ``eigenstates" like $|x\ra$ and $|p\ra$ 
for the continuous spectrum of some operator (like position and momentum here) are undefined in \Hs\ and hence have no inner product. Von Neumann circumvented this problem by first practically starting his book  with a tirade against Dirac's mathematics, and second, by writing down expressions like $\Tr(E(I)F(J))$, where $E(I)$ and $F(J)$ are the spectral projections for subsets $I$ and $J$ in the spectra of some self-adjoint operators $A$ and $B$, respectively. Unfortunately, these ``\tpies" are not only 
unnormalized; they may even be infinite. This was one of the reasons why von Neumann probably felt uncomfortable with his   formalism right from the start, and later sought a way out of this problem through a combination of lattice theory \`{a} la Birkhoff \& von Neumann (1936) and the theory of operator algebras he had also created himself, cf.\  \S\ref{OA}. The key is the existence of 
 type {\sc ii}$\mbox{}_1$ factors, which admit a \emph{normalised} trace $\mathrm{tr}$, i.e.\ $\mathrm{tr}(\mathbb{I})=1$.
Replacing $\Tr(E(I)F(J))$ by $\mathrm{tr}(E(I)F(J))$ then makes the \tpies\ finite as well as normalized,
but this only works if the spectral projections lie in the said factor. See Araki (1990) and R\'{e}dei (1996, 2001). \label{RFN}
}
\item[\emph{Ad 3(d)}.] Von Neumann's  Born probability to find a result $\lm\in I$ in a measurement of some self-adjoint operators $A$ in a state $\rh$ is given by $\Tr(\rh E(I))$, where $E(I)$ is the spectral projection for a subset $I$ in the spectrum of $A$ (and generalisations thereof to commuting observables). For $\rh=|\psi\ra\la\psi|$, $I=\{\lm\}$, and $E(I)=|\phv\ra\la\phv|$, assuming $A\phv=\lm\phv$ and the eigenvalue $\lm$ is nondegenerate, this recovers the transition probability in the previous item.
\item[\emph{Ad 3(e)}.] 
Von Neumann tried to \emph{prove} this identification by showing that if $\mathrm{Exp}$ is a \emph{linear} map from the (real) vector space of all bounded self-adjoint operators $A$ on some \Hs\ $H$
to $\R$ that is \emph{normalized} ($\mathrm{Exp}(\mathbb{I}))=1$), \emph{dispersion-free} ($\mathrm{Exp}(A^2)=\mathrm{Exp}(A)^2$), and satisfies a continuity condition (which is automatic if $H$ is finite-dimensional), then $\mathrm{Exp}(A)=\Tr (\rh A)$ for some density operator $\rh$ on $H$.
Unfortunately, he mistook this correct, non-circular, and interesting result for a proof that no hidden variables can underly \qm.
See Bub (2011) and Dieks (2016) for a fair and balanced account. 
\item[\emph{Ad 3(f)}.]  This was further developed by  Birkhoff \& von Neumann (1936), whose lattice-theoretic 
 calculus of such propositions initiated the field of quantum logic  (R\'{e}dei, 1998).
\end{trivlist}
\section{Quantum theory and functional analysis since 1932}
In this section we give a  brief overview of three areas of
 functional analysis that have had  fruitful interactions with quantum theory since the initial breakthrough during 1927--1932.
  \subsection{Unbounded operators}
  As already mentioned, motivated by \qm, von Neumann (1930ab, 1932) developed an abstract theory of self-adjoint operators, culminating in his spectral theorem. Some of this theory was constructed independently and simultaneously in the US by Stone (1932), who also found a result that von Neumann strangely missed and which is extremely important for the mathematical foundations of quantum theory:  \emph{Stone's Theorem} (to be distinguished from the closely related and equally famous \emph{Stone--von Neumann Theorem}),\footnote{Found independently by Stone (1931)  and von Neumann (1931), this theorem establishes the uniqueness of  irreducible \rep s of the canonical commutation relations that are integrable to unitary \rep s of  the Heisenberg group; the link between these notions was, prior to Stone's Theorem, first described by Weyl (1928).
See Summers (2001) and  Rosenberg (2004) for history and later developments.  } shows that (possibly unbounded) self-adjoint operators and (continuous) unitary \rep s of the additive group $\R$ on a \Hs\ are equivalent; this is a rigorous version of the link between a Hamiltonian $h$ and a unitary time-evolution $u_t=\exp(-ith)$.
 
The step from abstract theory to concrete examples that were actually useful for \qm\ turned out to be highly nontrivial.\footnote{Simon (2018, p.\ 176) mentions that around 1948 von Neumann told Bargmann that `\emph{self-adjointness for atomic Hamiltonians was an impossibly hard problem and that even for the Hydrogen atom, the problem was difficult and open}', adding that von Neumann's attitude may have discouraged work on the problem.} The first (and still most important) results in applying the abstract theory to atomic Hamiltonians are due to Kato (1951), who thereby established the study of \emph{Schr\"{o}dinger operators} as a mathematical discipline.\footnote{See also his textbook Kato (1966), followed by the  four-volume series Reed \& Simon (1972--1978),  many other books, and more briefly Simon (2000).  
 Kato's work is described in detail in Simon (2018).}
\subsection{Operator algebras and noncommutative geometry} \label{OA}
Von Neumann's contributions to the interplay between quantum theory and functional analysis did not end with his axiomatisations from the period 1927--1932. Parallel to (but apparently not inspired by) the development of quantum field theory,\footnote{In various places von Neumann mentioned \qm, ergodic theory, lattice theory, projective geometry (which he turned into \emph{continuous geometry}), and group representation theory as inspirations for operator algebras. Oddly, his direct attempts to describe quantum theory in a more algebraic fashion
 (Jordan, von Neumann, \& Wigner 1934; von Neumann, 1936), have  had little impact on physics so far.
}  in the 1930s he initiated the study of 
 \emph{rings of operators}, now called \emph{von Neumann algebras}. This theory was supplemented and refined by the work of Gelfand \& Naimark (1943), who (inspired by von Neumann algebras as well as by Gelfand's earlier work on commutative Banach algebras, but apparently not by physics directly) founded the field of \emph{\ca s}.\footnote{See  Petz \& R\'{e}dei (1995) for the history of von Neumann algebras, Doran \&\ Belfi (1986) for \ca s, and  Kadison (1982) for both. 
The founding papers `On rings of operators' {\sc i}--{\sc v} are collected in von Neumann (1961); nos.\ {\sc i}, {\sc ii}, and {\sc iv} are co-authored by von Neumann's assistant Murray. The paper Gelfand \& Naimark (1943) that started \ca s is reprinted (with notes by Kadison) in Doran (1994).} 

Jointly, von Neumann algebras and \ca s are called \emph{operator algebras}, which may be studied both abstractly and as algebras of concrete (bounded) operators on some \Hs. 
This flexibility allows a huge generalization of the pure \Hs\ formalism of von Neumann (1932), in which the algebra of all bounded operators on a given \Hs\ is replaced by an arbitrary (abstract) operator algebra. As such, one may continue to work with states, observables, and expectation values (Segal, 1947). 
From the 1960s onwards
operator algebras have become an important tool in mathematical physics, initially applied to quantum systems with infinitely many degrees of freedom,  as in quantum statistical mechanics (Ruelle, 1969; Bratteli \& Robinson, 1981, 1987; Haag, 1992; Simon, 1993) and quantum field theory (Haag, 1992; Araki, 1999;  Brunetti,  Dappiaggi, \& Fredenhagen, 2015).\footnote{The recollections of Haag (2010), arguably the main player in this field, are a valuable historical source.}  In turn, these physical applications have also given rise to various new mathematical ideas.\footnote{The work on the classification of von Neumann algebras for which Connes received the Fields Medal in 1982 is a good example: this relied on ideas originating in quantum statistical mechanics (Connes, 1994).}
 Furthermore, since the 1980s the field of operator algebras has been greatly refined and expanded by the toolkit of \emph{noncommutative geometry} (Connes, 1994), which so far has been applied to many areas of physics, ranging from  particle physics (Connes \& Marcolli, 2008; van Suijlekom, 2015) to deformation quantization and the classical limit of quantum mechanics (Rieffel, 1994; Landsman, 1998).
 \subsection{Distributions}
Another major development in functional analysis that is relevant for quantum physics was the theory of \emph{distributions}
due to Schwartz (1950--1951).\footnote{In Chapter VI of his autobiography, Schwartz (2001) gives some history. For example, on  pp.\ 227--228 he writes that he was inspired by the Dirac delta-function, PDEs,  divergent integrals, de Rham currents, and duality in topological vector spaces, but at the time of discovery (1944--1945) was unaware of previous relevant work by Heaviside, Bochner, Carleman, and Sobolev. See also Dieudonn\'{e} (1981), Chapter VIII.} Though closely related to Hilbert and Banach spaces through all kinds of natural embeddings and dualities, spaces of distributions belong to the wider class of \emph{locally convex topological vector spaces}, which incidentally were
 introduced by von Neumann (1935). The facts that the Dirac delta-function, which had annoyed von Neumann (1932, p.\ 2) so much, becomes a  well-defined object in distribution theory, and that the ``rigged Hilbert space" approach to distributions (Gelfand \& Vilenkin, 1964) even gives a rigorous and satisfactory version of Dirac's continuous eigenfunction expansions (Maurin, 1968),
 have had surprisingly little impact on \qm, even when it is seen in the light of mathematical physics.\footnote{See e.g.\ van Eijndhoven \& Graaf (1986) and Bohm (1994). The rigged \Hs\ approach  is not needed for spectral theory, though transition amplitudes like $\la x|p\ra$ in footnote \ref{RFN} now become well defined.
  } 
  
  Instead, the main applications of distribution theory to quantum physics have been to \emph{quantum field theory},
   in at least three originally different but closely related ways: 
 \begin{enumerate}
\item Through Wightman's \emph{Axiomatic Quantum Field Theory}, where quantum fields are  defined as unbounded operator-valued distributions (Streater \& Wightman, 1964).
\item Through \emph{causal perturbation theory},  an approach to renormalization based on the splitting of distributions with causal support into retarded and advanced parts.\footnote{This mathematically rigorous approach to renormalization has a long pedigree, but Epstein \& Glaser (1973) is generally regarded as a key paper.
For later work see e.g.\  Scharf (1995, 2001) and Rejzner (2016). }
\item Through \emph{microlocal analysis}, a phase space approach to distributions due to  
H\"{o}rmander (1990), which has become a key tool in quantum field theory on curved space-time.\footnote{See e.g.\ 
 B\"{a}r \&  Fredenhagen (2009), Brunetti, Dappiaggi, \& Fredenhagen (2015), and G\'{e}rard (2019).}
\end{enumerate}
 In fact, these three areas can no longer be separated, neither from each other nor from the operator-algebraic approach to quantum field theory mentioned in the previous subsection: their coalescence is one of the frontiers of contemporary research in mathematical physics.
\newpage
\section*{Epilogue}\addcontentsline{toc}{section}{Epilogue}
\begin{quote}\begin{small}
Functional analysis arose in the early twentieth century and gradually, conquering one stronghold after another, became a nearly universal mathematical doctrine, not merely a new area of mathematics, but a new mathematical world view. Its appearance was the inevitable consequence of the evolution of all of nineteenth-century mathematics, in particular classical analysis and mathematical physics. (\ldots) Its existence answered the question of how to state general principles of a broadly interpreted analysis in a way suitable for the most diverse situations. (Vershik, 2006, p.\ 438, quoted by MacCluer, 2009, p.\ vii).\end{small}
\end{quote}
This passage explains to some extent why the spectacular and unexpected application of functional analysis to quantum theory was possible:
though originating in problems from classical physics, the ``modernist" turn of mathematics towards abstraction and axiomatisation that brought the subject into the 20th century made almost every field of mathematics universally applicable. 
Moreover, much as quantum theory was originally meant to merely describe the atomic domain but subsequently, through its extension to quantum field theory  in fact turned out to be a theory of all of physics (except perhaps gravity), through von Neumann's invention of operator algebras as well as Schwartz's theory of distributions (both partly inspired by \qm),
 functional analysis continued to provide an appropriate mathematical language also for quantum field theory.

Having said this, the question \emph{why} functional analysis (here taken to be the original \emph{linear} theory) and especially \Hs s (or operator algebras) underlie quantum physics remains unanswered. Perhaps starting with Birkhoff \& von Neumann (1936), many people have tried to derive the mathematical formalism from plausible physical principles, but I believe that every such derivation so far contains a contingent or even incomprehensible part in order to derive the (complex) \Hs\ formalism.\footnote{In Birkhoff \& von Neumann (1936) the modular law  is already problematic; in refinements of their lattice-theoretic approach based on the reconstruction theorem of Sol\`{e}r (1995)
one has to assume orthomodularity \emph{and} the existence of an infinite orthonormal set (and still needs further arguments to single out $\C$ over $\R$ or $\mathbb{H}$), etc.
Mackey (1963)  himself admits  defeat by simply postulating that the lattice of propositions of a quantum system, for which he first gives many promising axioms, is isomorphic to the projection lattice $\CP(H)$ of some \Hs\ $H$. 
In my own approach based on the axiomatisation of transition probability spaces, axiom $C^*2$ on page 104 of Landsman (1998), which prescribes the transition probabilities of a 2-level system, seems to lack any physical justification.
See also Grinbaum (2007).}
 In this respect, the connection between quantum theory and functional analysis remains mysterious. 

Finally, let me note that  this was a \emph{winner's} (or ``whig") history, full of hero-worship:
 following in the footsteps of Hilbert, von Neumann established the link between quantum theory and functional analysis that has lasted. Moreover, partly through von Neumann's own contributions (which are on a par with those of Bohr, Einstein, and Schr\"{o}dinger), the precision that functional analysis has brought to quantum theory has greatly benefited the foundational debate. 
 However, it is simultaneously a \emph{loser's} history: starting with Dirac and continuing with Feynman, until the present day physicists have managed to bring quantum theory forward in utter (and, in my view, arrogant) disregard for the relevant mathematical literature.  As such, functional analysis has so far failed to make any real contribution  to quantum theory as a branch of physics (as opposed to mathematics), and in this respect its role seems to have been limited to something like classical music or other parts of human culture that adorn life but do not change the economy or save the planet. On the other hand, like General Relativity, perhaps the intellectual development reviewed in this paper is one of those human achievements that  make the planet worth saving. 
 \newpage
\addcontentsline{toc}{section}{References}
\begin{small}

\end{small}

\begin{thebibliography}{99}
\bibitem{} Araki, H. (1990). Some of the legacy of {J}ohn von {N}eumann in physics:
                  {T}heory of measurement, quantum logic and von {N}eumann algebras in physics.
                  In:  Glimm \emph{et al} (1990),  pp.\ 119--136.
\bibitem{} Araki, H. (1999). \emph{Mathematical Theory of Quantum Fields.} Oxford: Oxford University Press.  
\bibitem{}Banach, S. (1932).  \emph{Th\'{e}orie des Op\'{e}rations Lin\'{e}aires}. Warszawa:
Instytut Matematyczny Polskiej Akademii Nauk,  New York: Chelsea.
\bibitem{} B\"{a}r, C., Fredenhagen, K., eds. (2009). 
\emph{Quantum Field Theory on Curved Spacetimes: Concepts and Mathematical Foundations}.
Berlin: Springer. 
\bibitem{} Bernkopf, M.(1966). The development of function spaces with particular reference to their origins in integral equation theory.
\emph{Archive for History of Exact Sciences}  3, 1--96.
\bibitem{}Birkhoff, G. \&  Kreyszig, E. (1984). The establishment of functional analysis. \emph{Historia Mathematica} 11, 258--321.
\bibitem{}Birkhoff, G.,  von Neumann, J.  (1936). The logic of quantum mechanics.  
\emph{Annals of Mathematics}  37, 823--843.
\bibitem{} Blanchette, P. (2018).
 The Frege--Hilbert Controversy. \emph{The Stanford Encyclopedia of Philosophy (Fall 2018 Edition)}, Zalta, E.N. (ed.). \\\verb#https://plato.stanford.edu/archives/fall2018/entries/frege-hilbert/#.
 \bibitem{}Blumenthal, O. (1935). Lebensgeschichte. \emph{David Hilbert: Gesammelte Abhandlungen Vol. III}, pp.\ 388--429. Berlin: Springer.
 \bibitem{} Bohm, A. (1994). \emph{Quantum Mechanics: Foundations and Applications}. New York: 
  Springer.
 \bibitem{}Bratteli, O.,   Robinson, D.W. (1987). \emph{Operator
Algebras and Quantum Statistical Mechanics. Vol.\ I: C*- and
W*-Algebras, Symmetry Groups, Decomposition of States}. Second Edition.
 Berlin: Springer.
\bibitem{} Bratteli, O.,   Robinson, D.W. (1981). 
\emph{Operator Algebras and Quantum Statistical Mechanics. Vol.\ II:
Equilibrium States, Models in Statistical Mechanics}. Berlin: Springer.
\bibitem{} Br\'{o}dy, F., V\'{a}mos, eds.\ (1995).
\emph{The Neumann Compendium}.  Singapore: World Scientific. 
\bibitem{} Brunetti, R., Dappiaggi, C., Fredenhagen, K., eds.\ (2015). \emph{Advances in Algebraic Quantum Field Theory}.
Dordrecht: Springer.
\bibitem{} 
 Bub, J. (2011). Is von Neumann's `no hidden variables' proof silly?,
 \emph{Deep Beauty: Mathematical Innovation and the Search for Underlying Intelligibility in the Quantum World}, pp.\ 393--408. 
  Halvorson, H., ed.
   Cambridge: Cambridge University Press. 
\bibitem{}Connes, A. (1994). \emph{Noncommutative
Geometry}. San Diego: Academic Press. 
\bibitem{}Connes, A., Marcolli, M. (2008). \emph{Noncommutative
Geometry, Quantum Fields, and Motives}. New Delhi: Hindustan Book Agency. 
\bibitem{} Corry, L. (1997). David Hilbert and the axiomatization of physics (1894--1905). \emph{Archive for  History of  Exact Sciences} 51, 83--198. 
\bibitem{} Corry, L. (2004a). \emph{David Hilbert and the Axiomatization of Physics (1898--1918): From Grundlagen der Geometrie to Grundlagen der Physik}. Dordrecht: Kluwer. 
\bibitem{} Corry, L. (2004b). \emph{Modern Algebra and the Rise of Mathematical Structures. Second revised edition.}
Basel: Springer.
\bibitem{} Corry, L. (2018). Hilbert's sixth problem: between the foundations of geometry and the axiomatization of physics.
\emph{Philosophical Transactions of the Royal Society A}. DOI: \verb#10.1098/rsta.2017.0221#.
\bibitem{}   Courant, R., Hilbert, D. (1924).  \emph{Methoden der mathematischen Physik I}. Berlin: Springer.
 \bibitem{} Dieks, D. (2016). Von Neumann's impossibility proof: Mathematics
in the service of rhetorics. \emph{Studies in History and Philosophy of Modern Physics} 60, 136--148. 
\bibitem{}Dieudonn\'{e},  J. (1981). \emph{History of Functional Analysis}.  Amsterdam: North-Hol\-land. 
\bibitem{} Dijksterhuis, E.J. (1961). \emph{The Mechanisation of the World Picture}. Oxford: Oxford University Press. 
Translation of \emph{De Mechanisering van het Wereldbeeld} (Meulenhoff, Amsterdam, 1950). 
\bibitem{} Dirac, P.A.M. (1930). \emph{The Principles of Quantum Mechanics}. Oxford: Clarendon Press. 
 \bibitem{}Doran, R.S., ed.\ (1994).
\emph{C*-algebras: 1943--1993}. \emph{Contemporary  Mathematics} Vol.\ 167.
Providence: American Mathematical Society. 
\bibitem{}Doran, R.S., Belfi, V. (1986). \emph{Characterization of \ca s}. New York: Marcel Dekker.
\bibitem{} Dorier, J.-L. (1995). A general outline of the genesis of vector space theory. \emph{Historia Mathematica} 22, 227--261. 
\bibitem{} Duncan, A., Janssen, M. (2009). From canonical transformations to transformation theory, 1926--1927: The road to JordanÕs \emph{Neue Begr\"{u}ndung}. \emph{Studies in History and Philosophy of Modern Physics} 40, 352--362.
\bibitem{} Duncan, A., Janssen, M. (2013). (Never) Mind your $p$'s and $q$'s: Von Neumann versus Jordan on the foundations of quantum theory. \emph{The European Physical Journal H} 38, 175--259.
\bibitem{} Eijndhoven, S.J.L. van,  Graaf, J. de (1986). \emph{A Mathematical Introduction to Dirac's Formalism}.
Amsterdam: North-Holland (Elsevier).
\bibitem{}  Epstein, H., Glaser, V. (1973). The role of locality in perturbation theory. \emph{Annales de l'Institute Henri Poincar\'{e} A (Physique  th\'{e}orique)}  19,  211--295.
\bibitem{}  Ewald, W., Sieg, W. (2013). \emph{David Hilbert's Lectures on the Foundations of Arithmetic and Logic 1917-1933}. Heidelberg: Springer. 
  \bibitem{} Fischer, E. (1907). Sur la convergence en moyenne. \emph{Comptes Rendus de l'Acad\'{e}mie des Sciences}
144, 1022--1024. 
\bibitem{}  Fr\'{e}chet, N. (1906). Sur quelques points du calcul functionel. \emph{Rendiconti del Circolo Matematico di Palermo} 22, 1--74. 
\bibitem{} Gabriel, G., Kambartel, F., Thiel, C. (1980). \emph{Gottlob Frege's Briefwechsel mit D. Hilbert, E. Husserl, B. Russell, sowie ausgew\"{ahlte} Einzelbriefe Freges}. Hamburg: Felix Meiner Verlag.
\bibitem{} Gelfand, I.M., Naimark, M.A. (1943). On the imbedding of normed rings into the ring of operators in Hilbert space.
\emph{Sbornik: Mathematics} 12, 197--213.
\bibitem{} Gelfand, I.M.,  Vilenkin, N.J. (1964).
 \emph{Generalized Functions. Vol.\ 4: Some Applications of Harmonic Analysis. Rigged Hilbert Spaces.}
 New York: Academic Press.
 \bibitem{} G\'{e}rard, C. (2019). Microlocal analysis of quantum fields on curved spacetimes. \verb#https://arxiv.org/abs/1901.10175#.
 \bibitem{} Glimm, J., Impagliazzo, J., Singer, I., eds. (1990). \emph{The Legacy of John von Neumann}.
 \emph{Proceedings of Symposia in Pure Mathematics} 50. 
 Providence: American Mathematical Society.
\bibitem{} Gray, J.D. (2008). \emph{Plato's Ghost: The Modernist Transformation of Mathematics}. Princeton: Princeton University Press. 
\bibitem{}  Grinbaum, A. (2007). Reconstruction of quantum theory. \emph{British Journal for the Philosophy of Science} 58,  387--408.
 \bibitem{}  Haag, R. (1992).   \emph{Local Quantum Physics: Fields, Particles, Algebras}. Heidelberg: Springer.
\bibitem{}  Haag, R. (2010).  Some people and some problems met in half
a century of commitment to mathematical physics. \emph{The European Physics Journal H} 35, 263--307. 
\bibitem{}Heims, S.J. (1980). \emph{John von Neumann and Norbert Wiener: From Mathematics to the Technologies of Life and Death}. Cambridge (Mass.): MIT Press. 
\bibitem{}  Hilbert, D. (1902). Mathematical Problems. Lecture delivered before the International Congress of Mathematicians at Paris in 1900. \emph{Bulletin of the American Mathematical Society} 8, 437--479. Translated from  \emph{G\"{o}ttinger Nachrichten}, 1900, pp. 253--297.
\bibitem{}  Hilbert, D. (1906). Grundz\"{u}ge einer allgemeinen Theorie der linearen Integralgleichungen. Vierte Mitteilung.
\emph{Nachrichten von der Gesellschaft der Wissenschaften zu G\"{o}ttingen, Mathematisch-Physikalische Klasse},  157--227. Reprinted in Hilbert (1912).
\bibitem{}  Hilbert, D. (1912). \emph{Grundz\"{u}ge einer allgemeinen Theorie der linearen Integralgleichungen}.
Leipiz und Berlin: Teubner. \verb#https://archive.org/details/grundzugeallg00hilbrich#.
\bibitem{}  Hilbert, D. (1918). Axiomatisches Denken. \emph{Mathematische Annalen} 78, 405--415. Reprinted in \emph{David Hilbert: Gesammelte Abhandlungen Vol. III}, pp.\ 146--156 (1935).  Berlin: Springer.
\bibitem{}  Hilbert, D., Bernays, P.  (1934, 1939). \emph{Grundlagen der Mathematik, Bd.\ I, II}. Berlin: Springer.  
\bibitem{}  Hilbert, D., von Neumann, J., Nordheim, L. (1927). \"{U}ber die Grundlagen der Quantenmechanik.
\emph{Mathematische Annalen} 98, 1--30.
\bibitem{} H\"{o}rmander, L. (1990). \emph{The Analysis of Linear Partial Differential Operators I. Second Edition.} Berlin: Springer.
\bibitem{} Jammer, M. (1989). \emph{The Conceptual Development of Quantum Mechanics. Second Edition}. 
New York: American Institute of Physics.
\bibitem{} Jordan, P., von Neumann, J., Wigner, E.P. (1934). 
On an algebraic generalization of the quantum mechanical formalism. 
\emph{Annals of Mathematics 35, 29--64.}
\bibitem{}
Kadison, R.V.  (1982). Operator algebras: The first forty years. \textit{Proceedings of  Symposia in Pure Mathematics} 
38(1), pp.\ 1--18. Providence: American Mathematical Society. 
\bibitem{}  Kato, T. (1951). Fundamental properties of Hamiltonian operators of Schr\"{o}dinger type.
\emph{Transactions of the American Mathematics Society} 70, 195--211.
\bibitem{}  Kato, T. (1966). \emph{Perturbation Theory for Linear Operators}. Berlin: Springer. 
\bibitem{}  Lacki, J. (2000). The early axiomatizations of quantum mechanics: Jordan, von Neumann and the continuation of Hilbert's program. \emph{Archive for History of Exact Sciences} 54, 279--318. 
 \bibitem{}  Landsman, K. (1998). \emph{Mathematical Topics Between Classical and Quantum Mechanics}. New York: Springer. 
\bibitem{}   Landsman, K. (2017). \emph{Foundations of Quantum Theory: From Classical Concepts to Operator Algebras}. 
Cham: Springer Open.  \verb#https://www.springer.com/gp/book/9783319517766#.
    \bibitem{} MacCluer, B.D. (2009). \emph{Elementary Functional Analysis}. New York: Springer.  
          \bibitem{} Mackey, G.W. (1963). \emph{The Mathematical Foundations of Quantum Mechanics}. New York: Benjamin.  
      \bibitem{} Macrae, N. (1992). \emph{John von Neumann: The Scientific Genius Who Pioneered the Modern Computer, Game Theory, Nuclear Deterrence, and Much More}. Providence: American Mathematical Society. 
              \bibitem{} Majer, U. (2014). The ``axiomatic method" and its constitutive role in physics.  
        \emph{Perspectives on Science} 22, 56--79.
            \bibitem{}   Majer, U., Sauer, T., Eds.\ (2021).  \emph{David Hilbert's Lectures on the Foundations of Physics 1898--1914}.     
        \bibitem{}  Mancosu, P. (2003). The Russellian influence on Hilbert and his school. \emph{Synthese} 137, 59--101.
          \bibitem{}   Maurin, K. (1968). \emph{Generalized Eigenfunction Expansions and Unitary Representations of Topological Groups}. Warsaw: Polish Scientific Publishers. 
\bibitem{}  Mehra, J.,  Rechenberg, H. (2000). \emph{The Historical Development of Quantum Theory. Vol.\ 6:
The Completion of Quantum Mechanics 1926--1941.    Part 1: The Probabilistic Interpretation and the Empirical and Mathematical Foundation of Quantum Mechanics, 1926-1936}.  New York: Springer-Verlag.
\bibitem{} Mehrtens, H. (1990). \emph{Moderne Sprache Mathematik}. Frankfurt a/M: Suhrkamp.
\bibitem{}Monna, A.F. (1973). \emph{Functional Analysis in Historical Perspective}. Utrecht: Oosthoek.
\bibitem{}Monna, A.F. (1973). \emph{Dirichlet's Principle: A mathematical Comedy of Errors and its Influence on the Development of Analysis.}  Utrecht: Oosthoek, Scheltema and Holkema. 
\bibitem{} Moore, G.H. (1995). The axiomatization of linear algebra: 1875--1940.  \emph{Historia Mathematica} 22, 262--303. 
\bibitem{}Neumann, J. von (1927a). Mathematische Begr\"{u}ndung  der Quantenmechanik. 
\emph{Nach\-richten von der Gesellschaft der Wissenschaften zu G\"{o}ttingen, Mathematisch-Physik\-alische Klasse},
1--57.  \verb#https://eudml.org/doc/59215#.
\bibitem{}Neumann, J. von (1927b). Wahrscheinlichkeitstheoretischer Aufbau der Quantenmechanik.
\emph{ibid.}
245--272.     \verb#https://eudml.org/doc/59230#.
 \bibitem{}Neumann, J. von (1930a). Allgemeine Eigenwerttheorie hermitischer Funktionaloperatoren.
   \emph{Mathematische Annalen} 102, 49--131. 
 \bibitem{}Neumann, J. von (1930b). Zur Algebra der Funktionaloperationen und  Theorie der normalen Operatoren.
  \emph{Mathematische Annalen} 102, 370--427. 
\bibitem{}Neumann, J. von (1931). Die Eindeutigkeit der Schr\"{o}deringerschen Operatoren.
   \emph{Mathematische Annalen} 104, 570--578.
\bibitem{}Neumann, J. von (1932). \emph{Mathematische Grundlagen der Quantenmechanik.}
Berlin: Springer--Verlag. English translation: \emph{Mathematical Foundations of Quantum Mechanics}. Princeton: Princeton University Press (1955).
\bibitem{}Neumann, J. von (1935). On complete topological spaces. \emph{Transactions of the  American Mathematica Society}  37, 1--20.
 \bibitem{}  Neumann, J. von (1936).  On an algebraic generalization of the quantum mechanical formalism (Part {\sc i}).
 \emph{Sbornik: Mathematics (N.S.)} 1, 415--484.
\bibitem{} Neumann, J. von (1961).
\emph{Collected Works, Vol.\ III: Rings of Operators}.
Taub, A.H., ed. Oxford: Pergamon Press.
\bibitem{}Oxtoby, J.C., Pettis, B.J., Price, G.B. (1958). John von Neumann: 1903--1957.
\emph{Bulletin of the American Mathematical Society} 64, No.\ 3, Part 2. 
\bibitem{} Peano, G. (1888). \emph{Calcolo geometrico secondo l'Ausdehnungslehre di H. Grassmann: preceduto dalla operazioni della logica deduttiva}. Torino: Fratelli Bocca Editori. 
\bibitem{} Petz, D., R\'{e}dei, M. (1995). John von Neumann and the theory of operator algebras.
\emph{The Neumann Compendium}, pp.\ 163--181. Br\'{o}dy, F., V\'{a}mos, eds.\ Singapore: World Scientific. 
\bibitem{}Pier, J. (2001). \emph{Mathematical Analysis During the 20th Century}. New York: Oxford University Press.
\bibitem{} Pietsch, W. (2007). \emph{History of Banach Spaces and Linear Operators}. Basel:  Birkh\"{a}user.
\bibitem{} Pyenson, L.R. (1982). Relativity in late-Wilhelminian Germany: The appeal to a preestablished harmony between mathematics and physics. \emph{Archive for History of Exact Sciences} 27, 137-155.
\bibitem{}R\'{e}dei, M. (1996). Why John von Neumann did not like the \Hs\ formalism of \qm\ (and what he liked instead).
\emph{Studies in History and Philosophy of Modern Physics} 27, 493--510. 
\bibitem{}R\'{e}dei, M. (1998). \emph{Quantum Logic in Algebraic Approach}. Dordrecht: Kluwer Academic Publishers.
\bibitem{}R\'{e}dei, M. (2001). Von Neumann's concept of quantum logic and quantum probability. 
In:   R\'{e}dei \& St\"{o}ltzner (2001), pp.\ 153--172.
\bibitem{}R\'{e}dei, M. (2005a). John von Neumann on mathematical and axiomatic physics. \emph{The Role of Mathematics in the Physical Sciences},  pp.\ 43--54. Boniolo, G.\ et al, eds.  Dordrecht: Springer. 
\bibitem{}R\'{e}dei, M. (2005b). \emph{John von Neumann: Selected Letters}. 
 Providence: American Mathemematical Society. 
\bibitem{}  R\'{e}dei, M., St\"{o}ltzner, eds.\ (2001).   \emph{John von Neumann and the Foundations of Quantum Physics}.
   Dordrecht: Kluwer. 
   \bibitem{}  R\'{e}dei, M., St\"{o}ltzner (2006), Soft axiomatisation: John von Neumann on method and von Neumann's method in the physical sciences. \emph{Intuition and the Axiomatic Method}, Carson, E., Huber, R. (eds.), pp.\  235--250. Dordrecht: Springer.
\bibitem{} Reed, M.,  Simon, B. (1972--1978).  \emph{Methods of Modern Mathematical Physics. Volume I: Functional Analysis.
Volume II: Fourier Analysis, Self-Adjointness. Volume III: Scattering Theory. 
Volume IV:  Analysis of Operators.}
  New York: Academic Press.
   \bibitem{} Reid, C. (1970). \emph{Hilbert}. Berlin: Springer. 
      \bibitem{} Rejzner, K. (2016). \emph{Perturbative Algebraic Quantum Field Theory:
An Introduction for Mathematicians}. Cham: Springer. 
\bibitem{} Rieffel, M.A. (1994). 
  Quantization and $C\sp *$-algebras.   \emph{Contemporary Mathematics} 167,  66--97.
  \bibitem{}Riesz, F. (1907). Sur les syst\`{e}mes orthogonaux des functions. \emph{Comptes Rendus de l'Acad\'{e}mie des Sciences}
144, 615--619.
  \bibitem{}Riesz, F. (1909). Untersuchungen \"{u}ber Systeme integrierbarer Funktionen. \emph{Mathematische Annalen} 
  69, 449--497.
\bibitem{}Riesz, F. (1913). \emph{Les Syst\`{e}mes d'\'{E}quations Lin\'{e}aires \`{a} une Infinit\'{e} d'Inconnues}.
Paris: Gauthiers--Villars.
\bibitem{}Riesz, F. (1918). \"{U}ber lineare Funktionalgleichungen. \emph{Acta Mathematica} 41, 71--98.
\bibitem{} Rosenberg, J. (2004). A selective history of the Stone--von Neumann Theorem. \emph{Contemporary Mathematics}
365, 331Ð353.
\bibitem{} Rowe, D.E. (2018). \emph{A Richer Picture of Mathematics: The G\"{o}ttingen Tradition and Beyond}. Cham: Springer. 
\bibitem{} Ruelle, D. (1969). \emph{Statistical Mechanics: Rigorous Results}. New York: Benjamin.
\bibitem{} Sauer, T., Majer, U., Eds. (2009). \emph{David Hilbert's Lectures on the Foundations of Physics 1915--1927}.
Dordrecht: Springer. 
 \bibitem{} Scharf, G. (1995). \emph{Finite Electrodynamics: The Causal Approach. Second Edition}.
 Berlin: Springer. 
  \bibitem{} Scharf, G. (2001). \emph{Quantum Gauge Theories: A True Ghost Story}.
  New York: Wiley.
\bibitem{} Schirrmacher, A. (2019). \emph{Establishing Quantum Physics in G\"{o}ttingen: David Hilbert, Max Born, and Peter Debye in Context, 1900--1926}. Cham: Springer. 
\bibitem{} Schmidt, E. (1908). \"{U}ber die Aufl\"{o}sung linearer Gleichungen mit unendlich vielen Unbekannten.
\emph{Rendiconti del Circolo Matematico di Palermo} {\sc xxv}, 53--77. 
\bibitem{} Sch\"{o}nflies, A. (1908). Die Entwicklung der Lehre von den Punktmannigfaltigkeiten, Zweiter Teil.
\emph{Jahresberichte der deutschen Mathematiker-Vereinigung, Erg\"{a}nzungsband}.  Leipzig: Teubner.
\bibitem{} Schwartz, L. (1950--1951). \emph{Th\'{e}orie des Distributions. Vols. 1 \& 2}. Paris: Hermann. 
\bibitem{} Schwartz, L. (2001). \emph{A Mathematician Grappling with his Century}. Basel: Birkh\"{a}user. 
 \bibitem{} Segal, I.E. (1947). Postulates for general \qm. \emph{Annals of Mathematics} 48, 930--948. 
\bibitem{}Sieg, W. (2013). \emph{Hilbert's Programs and Beyond}. Oxford: Oxford University Press.
\bibitem{}Siegmund-Schultze, R. (1982). Die Anf\"{a}nge der Funktionalanalysis und ihr Platz im Unwandlungsproze\ss\ der Mathematik um 1900. \emph{Archive for History of Exact Sciences} 26, 13--71.
\bibitem{}Siegmund-Schultze, R. (2003). The origins of functional analysis. \emph{A History of Analysis},  pp.\ 385--408.
Jahnke, H. N., ed.  Providence: American Mathematical Society. 
 \bibitem{}   Simon, B. (1993). \emph{The Statistical Mechanics of Lattice Gases. Vol.\ I}. Princeton: Princeton University Press. 
\bibitem{} Simon, B.  (2000). Schr\"{o}dinger operators in the twentieth century.
\emph{Journal of Mathematical Physics}  41, 3523--3555. 
\bibitem{} Simon, B. (2018). Tosio Kato's work on non-relativistic quantum mechanics, part 1 and part 2. \emph{Bulletin of Mathematical Sciences} 8, 121-232 and \emph{ibid.} 9, 1--99. 
\bibitem{}  Sol\`{e}r, M.P. (1995). Characterization of Hilbert spaces by orthomodular spaces. \emph{Communications in Algebra} 23, 219--243
\bibitem{}Steen, L.A. (1973). Highlights in the history of spectral theory. \emph{American Mathematical Monthly} 80, 359--381.
\bibitem{}Stone, M.H. (1931). Linear transformations in \Hs, III: Operational methods and group theory.
\emph{Proceedings of the National Academy of Sciences of the United States of America} 16, 172--175. 
\bibitem{}Stone, M.H. (1932). \emph{Linear Transformations in \Hs\ and their Applications to Analysis.}  Providence: American Mathematical Society.
\bibitem{}  St\"{o}ltzner, M. (2001). Opportunistic axiomatics--von Neumann and the methodology of mathematical physics.
In:   R\'{e}dei \& St\"{o}ltzner (2001), pp.\ 35--62.
\bibitem{}Streater, R.F., Wightman, A.S. (1964). \emph{PCT, Spin and Statistics, and All That}. New York: Benjamin. 
\bibitem{} Suijlekom, W.D. van (2015). \emph{Noncommutative Geometry and Particle Physics}. Dordrecht: Springer. 
\bibitem{} Summers, S.J. (2001). On the Stone--von Neumann uniqueness theorem and its ramifications.
  \emph{John von Neumann and the Foundations of Quantum Physics}, pp.\ 135--152.
   R\'{e}dei, M., St\"{o}ltzner, eds.\  Dordrecht: Kluwer. 
\bibitem{} Vershik, A.M. (2006). The life and fate of functional analysis in the twentieth century.
\emph{Mathematical Events of the Twentieth Century}, eds.\   Bolibruch, A.A. {\em et al},  pp. 437--447. Berlin: Springer.
\bibitem{} Volkert, K. (2015). \emph{David Hilbert: Grundlagen der Geometrie}. Berlin: Springer.
 \bibitem{}Vonneumann, N. (1987). \emph{John von Neumann as Seen by his Brother}. Typescript. 
\bibitem{} Weyl, H. (1918). \emph{Raum - Zeit - Materie}. Berlin: Springer. 
\bibitem{} Weyl, H. (1928). \emph{Gruppentheorie und Quantenmechanik}. Leipzig: Hirzel.
\bibitem{} Weyl, H. (1951). A half-century of mathematics. \emph{The American Mathematical Monthly} 58, 523--553. 
\end{thebibliography}
\end{document}